\documentclass[twocolumn,showpacs,preprintnumbers,amsmath,amssymb]{revtex4}

\usepackage{epsfig}
\usepackage{dcolumn}
\usepackage{bm}

\def\ee{\end{equation}}
\def\be{\begin{equation}}
\def\ba{\begin{eqnarray}}
\def\ea{\end{eqnarray}}
\def\ll{\label}

\begin{document}

\preprint{APS/123-QED}
\title{Statistical Mechanics in the Extended Gaussian Ensemble}
\author{Ramandeep S. Johal}   
\email{rjohal@ecm.ub.es}   
\author{Antoni Planes}
\email{toni@ecm.ub.es}
\author{Eduard Vives}  
\email{eduard@ecm.ub.es}  \affiliation{   Departament  d'Estructura  i
Constituents  de la  Mat\`eria, Universitat  de Barcelona  \\ Diagonal
647, Facultat de F\'{\i}sica, 08028 Barcelona, Catalonia}
\date{\today}

\begin{abstract}
The extended gaussian ensemble (EGE) is introduced as a generalization
of the canonical ensemble. The  new ensemble is a further extension of
the Gaussian ensemble  introduced by J. H. Hetherington  [J. Low Temp.
Phys.  {\bf 66}, 145 (1987)].  The statistical mechanical formalism is
derived  both from the  analysis of  the system  attached to  a finite
reservoir  and from  the Maximum  Statistical Entropy  Principle.  The
probability of  each microstate depends on two  parameters $\beta$ and
$\gamma$ which  allow to  fix, independently, the  mean energy  of the
system  and the  energy fluctuations  respectively.  We  establish the
Legendre  transform   structure  for  the   generalized  thermodynamic
potential and propose a stability  criterion.  We also compare the EGE
probability distribution with the $q$-exponential distribution.  As an
example,  an application  to a  system with  few independent  spins is
presented.
\end{abstract}


\maketitle
%
\section{Introduction}
\label{Intro}
%
The development  of statistical mechanics based on  ensemble theory is
founded on  the postulate of  ``equal {\it a  priori} probabilities'',
which is assumed to apply to all microstates consistent with the given
macrostate   of   an    isolated   system   \cite{Pathria1996}.    The
corresponding  statistical ensemble  is  the so-called  microcanonical
ensemble.   A   representative  system   in  this  ensemble   has  all
``mechanical'' variables such as energy $E$, volume $V$, magnetization
$M$ etc., fixed.  For convenience in calculations, other ensembles are
used which invariably suppose the  existence of a subsidiary system or
reservoir in  contact with  the actual system.   For instance,  in the
canonical  ensemble the  walls of  the  system permit  an exchange  of
energy with the reservoir while  in the grand canonical ensemble, both
energy  and  matter  can  be  exchanged.  In  general,  the  different
ensembles are constructed by allowing one or more mechanical variables
to fluctuate.  The  exchange of each of these  variables is controlled
by  a parameter  which  is  a characteristic  of  the reservoir.   For
instance, in  the case  of the canonical  ensemble, this  parameter is
precisely  the temperature of  the reservoir  and determines  the mean
energy of the  system.  Actually, this is adequate  when the reservoir
is a very large system  that can exchange arbitrary amounts of energy,
without  modification  of  its  intensive  properties.   In  practical
situations, this  is not always  the case.  However, very  few studies
have been  devoted to analyse the consequences  of possible deviations
from these ideal reservoir properties.

In this paper,  we develop the statistical mechanics  of a system that
can  exchange energy  with  a finite  reservoir  characterized by  two
parameters:   $\beta$   and   $\gamma$.   These   parameters   control
independently  the   mean  energy  of   the  system  and   its  energy
fluctuations,  respectively.   The  corresponding statistical  ensemble
represents  a generalization  of the  canonical ensemble  and  will be
called the  extended gaussian ensemble (EGE).  A  similar ensemble was
already  developed, in  a more  restricted framework,  by Hetherington
\cite{Hetherington1987}.  The author considered that the sample system
was in contact with a finite reservoir with size dependent properties.
The  so-called  gaussian  ensemble   was  introduced  so  that  it  is
equivalent to  the canonical ensemble  in the limit of  large systems,
except in the energy range of a first-order transition. Interestingly,
it enables  a smooth interpolation between the  microcanonical and the
canonical ensembles.   Taking into account these  features, Challa and
Hetherington  \cite{Challa1988a,Challa1988b}  showed  the interest  of
this ensemble for Monte Carlo simulation studies of phase transitions.
They demonstrated  a significant reduction in  computer time (compared
to standard  simulations in the canonical ensemble)  and, its adequacy
for   distinguishing   second-order   from  first-order   transitions.
Compared  to  the  EGE  introduced  in the  present  paper,  the  main
difference  arises from  the fact  that in  the gaussian  ensemble the
sample and  the reservoir are assumed to  be statistically independent
which implies the additivity  of the corresponding entropies.  This is
not assumed in our formalism.  The consequences are important and will
be discussed in depth in this work.

The  present formalism  can be  considered  as an  alternative to  the
statistical  mechanics based  on  non-additive generalized  entropies.
Actually the study  of such generalized entropies has  generated a lot
of interest  in the  past fiveteen years.   The motivation for  the so
called Tsallis  statistical mechanics has been to  extend the standard
Boltzmann-Gibbs    framework   to   include    non-extensive   systems
\cite{Tsallis1988}.   Among  different  interpretations, it  has  been
suggested that Tsallis formalism corresponds to an ensemble describing
a      system      attached       to      a      finite      reservoir
\cite{Almeida2001,Plastino1994}.   Although a  large number  of papers
have been published, the physical meaning of many related issues is 
still open
to  discussion  \cite{Nauenberg2003,Tsallis2003}.   The EGE  formalism
that  we  propose in  this  paper,  provides  a clear  and  consistent
framework for the statistical mechanics with a finite reservoir.
 
The  paper is  organized  as follows:  in  sections \ref{Contact}  and
\ref{Maxent}, the EGE is founded from the analysis of a contact with a
finite reservoir  and from  the Maximum Statistical  Entropy Principle
respectively.   In   section  \ref{Thermo},  the   main  thermodynamic
relations  are  derived. In  section  \ref{nonadd},  we highlight  the
non-additive  nature  of  the  thermodynamic  formalism.   In  section
\ref{Stabi},   a  stability   criterion  is   proposed.    In  section
\ref{Qexp}, the equilibrium distributions of the EGE are compared with
the $q$-exponential distributions.  In section \ref{Appli}, we present
an example of application to  a system of independent spins.  Finally,
in section \ref{Concl} we summarize and conclude.
%
\section{Contact with a Finite Reservoir}
\label{Contact}
%
Let us consider a system (that  we will call the sample) in contact with a
reservoir.  Let us call the energy  of the sample $E_1$ and the energy
of the reservoir  $E_2$.  The sample and the  reservoir together form
an isolated system so  that  $E=E_1+E_2$  is  constant.   Let  us  also  define
$\Omega_{2}(E_{2})$   as   the    number   of   microstates   of   the
reservoir.  Following Callen  \cite{Callen1985}, the  probability that
the system 1 is in a certain microstate with energy $E_1$ is given by
\begin{equation}
p_1(E_1)=\frac{\Omega_2(E-E_1)}{\Omega_{1+2}(E)},
\label{firsteq}
\end{equation}
where $\Omega_{1+2}(E)$  is the total  number of states  available for
the set $1+2$  (Note that we do not assume  that $\Omega_{1+2}$ can be
factorized as a  product $\Omega_{1} \Omega_{2}$ ). Let  us define the
entropy of the reservoir as:
\begin{equation}
S_2(E_2) = \ln \Omega_2(E_2), 
\ll{s2log}
\end{equation}
(all along  the paper  we choose the  entropy units so  that Boltzmann's
constant, $k_B=1$.)
Therefore:
\begin{equation}
p_1(E_1)=\frac{e^{S_2(E-E_1)}}{\Omega_{1+2}(E)}.
\label{prob1}
\end{equation}
The energy of  the sample will, in general fluctuate.  Let us call $U$
its mean (equilibrium) value.   We can develop $S_2(E-E_1)$ around the
equilibrium value $E-U$ as:
\begin{eqnarray}
S_2(E-E_1) & =& S_2(E-U) +  \left . \frac{d S_2}{d E_2} \right |_{E-U}
 (U-E_1) \nonumber \\ &+& \frac{1}{2!} \left . \frac{d^2 S_2}{d E_2^2}
 \right |_{E-U} (U-E_1)^2 + {\cal O}(U-E_1)^3.  \nonumber \\ & &
\label{expands2}
\end{eqnarray}
The  derivatives  in  the  right  hand side  of  this  expression  are
quantities which depend only on the reservoir. We define
\begin{equation}
\left . \frac{d S_2}{d E_2} \right |_{E-U} = \beta,
\label{defbeta}
\end{equation}
and 
\begin{equation}
\left . \frac{d^2 S_2}{d E_2^2}\right |_{E-U} = -2 \gamma.
\label{defgam}
\end{equation}
The standard  canonical ensemble is characterized by  a reservoir with
constant $\beta$  (independent of $E_2$),  which implies $\gamma  = 0$
and  there is  no term  beyond the  first order  term in  the 
Eq.~(\ref{expands2}).  However,  in the present paper, we  consider a more
general reservoir  for which $\gamma \neq  0$.  Thus in  this EGE, the
reservoir  is characterized  by  the pair  of  parameters $\beta$  and
$\gamma$.   The  thermodynamic meaning  of  these  parameters will  be
clarified  in the  following  sections.  To  explicitly highlight  the
effects of this  modification and also for the  sake of simplicity, we
assume  that   the  cubic  and   higher  order  terms   vanish.   Then
substituting  (\ref{expands2})  in   (\ref{prob1})  and  denoting  the
energies of  the microstates of the sample  by $\epsilon_i$ ($i=1,...,
M$), we obtain
\begin{equation}
p_i = \frac{1}{Z_G}{\exp} [ -\beta \epsilon_i - \gamma (\epsilon_i -U)^2 ],
\label{piexp}
\end{equation}
where the normalization constant $Z_G$ is given by
\begin{equation}
Z_G    =   \sum_{i=1}^{M}    {\exp}[    -\beta\epsilon_i   -    \gamma
(\epsilon_i-U)^2].
\label{defzg}
\end{equation}
The   subscript   $G$  only indicates   the   ``gaussian''   form  of   the
probabilities.  Note that $U$ is  the mean energy and must be obtained
self-consistently from the equation:
\begin{equation}
U Z_G  = \sum_{i=1}^{M}  \epsilon_i {\exp}[ -\beta\epsilon_i  - \gamma
(\epsilon_i-U)^2].
\label{defu}
\end{equation}
Eqs.~(\ref{piexp}),  (\ref{defzg})  and  (\ref{defu})  reduce  to  the
standard canonical ensemble  definitions when $\gamma=0$. Therefore, it
is natural  to relate the parameter  $\gamma$ with the  finite size of
the reservoir.
%
\section{Maximum Statistical Entropy Principle}
\label{Maxent}
%
In this  section, we derive  the probability law  of Eq.~(\ref{piexp})
from different arguments.  This leads to a better understanding of the
parameters  $\beta$  and  $\gamma$  as parameters  characterizing  the
``equilibrium''  distribution of  the sample.  To  derive the
probability   distribution  from   the  Maximum   Statistical  Entropy
Principle,  we maximize  the standard  Gibbs-Boltzmann-Shannon entropy
given by
\begin{equation}
S_G = -\sum_{i=1}^{M} p_i \ln p_i,
\label{defsg}
\end{equation}
subject to the constraints of normalisation of the probability,
the given mean value of the energy and the  fixed value of the
fluctuations, respectively  as
\begin{eqnarray}
\sum_{i=1}^{M}  p_i &  =  & 1,  \\  
\langle \epsilon_i  \rangle \equiv  \sum_{i=1}^{M} \epsilon_i  p_i &  = &  
U, \\  
\langle (\epsilon_i -U)^2 \rangle \equiv \sum_{i=1}^{M} (\epsilon_i  -U)^2 
p_i  & =  & W.
\label{constraints}
\end{eqnarray}  
Then the  maximization procedure is  done by introducing  the Lagrange
multipiers  $\lambda$,   $\beta$  and  $\gamma$   for  the  respective
constraints, and maximizing the following functional ${\cal L}$:
\ba
{\cal L} &=& -\sum_i p_i \ln p_i  
- \lambda \left( \sum_i p_i -1 \right) - \beta 
\left( \sum_i \epsilon_i  p_i - U \right)
\nonumber \\
& & - \gamma \left( \sum_i (\epsilon_i -U)^2 p_i - W \right).
\ea
By requiring the condition:
\begin{equation}
\frac {\partial {\cal L}}{\partial p_i} = 0, 
\end{equation}
it  is  easy  to  see   that  the  optimum  form  of  the  probability
distribution  is   given  by  the   expression  in  Eq.~(\ref{piexp}).
Therefore  $\beta$  and  $\gamma$,  within this  context,  are  simply
Lagrange  multipliers that  allow  to fix,  self-consistently, a  mean
value of the energy $U=\langle  \epsilon_i \rangle$ and a specific value of the
variance $W=\langle (\epsilon_i-U)^2 \rangle $.
%
\section{Thermodynamic Relations}
\label{Thermo}
%
We define a thermodynamic potential $\Phi (\beta, \gamma)$ as
\begin{equation}
\Phi (\beta, \gamma)= \ln Z_G.
\label{defphi}
\end{equation}
By  differentiating  Eq.~(\ref{defzg}),   it  can  be  straightforwardly
obtained that:
\begin{eqnarray}
- \left( \frac{\partial \Phi}{\partial \beta} \right )_{\gamma} & = &
U(\beta,  \gamma)   ,  \label{dphib}  \\ 
- \left(  \frac{\partial \Phi}{\partial  \gamma}\right)_{\beta} &  = &
W(\beta,\gamma). 
\label{dphigam}
\end{eqnarray}
The second derivative renders:
\begin{equation}
- \left( \frac{\partial^2 \Phi}{\partial \beta^2} \right)_{\gamma}
 =  - \left( \frac{\partial U}{\partial \beta} \right)_{\gamma}
=   \frac{1}{  W^{-1}(\beta,\gamma) - 2{\gamma}},
\label{d2phib} 
\end{equation}
which represents  a generalization of the  standard formula for
energy fluctuations in the  canonical  ensemble.   It is  natural  
to define  the extended heat capacity as:
\begin{equation}
{\cal  C} \equiv  - \beta^2  \left( \frac{\partial  U}{\partial \beta}
\right)_{\gamma}= \frac{\beta^2 W}{1-2\gamma W}.
\label{defc}
\end{equation}
This   equation   is   the   same   that  was   already   derived   in
Ref.~\onlinecite{Challa1988b}. Note  that, contrary to  what happens
in the standard canonical ensemble, the positivity of the fluctuations
$W$ does not guarantee the positivity of $\cal C$.

For $\gamma  \to 0$, it is  seen that the  relations (\ref{dphib}) and
(\ref{d2phib})  go to  the  corresponding relations  for  the case  of
canonical ensemble.   Also in  this limit, from  Eqs.  (\ref{dphigam})
and (\ref{d2phib}) we get an interesting relation given by:
\begin{equation}  
\lim_{\gamma  \to  0}  \left( \frac{\partial  \Phi}{\partial  \gamma}  \right)
_{\beta}= \left( \frac{\partial^2 \Phi}{\partial \beta^2}
\right)_{\gamma}, \ll{diffueqn}
\end{equation} 
which  resembles  in form with a diffusion equation.

The entropy  $S_G$ as given  by (\ref{defsg}) is the  inverse Legendre
transform of $\Phi(\beta, \gamma)$, and can be expressed as:
\begin{equation}
S_G(U,W) = \beta U + \gamma W + \Phi, \ll{legend}
\end{equation}
whereby $S_G$ is a function of the specified values of the constraints
i.e.  $U$  and $W$.  Therefore  we  have  the following  thermodynamic
relations
\begin{eqnarray}
\left( \frac{\partial S_G}{\partial U} \right)_{W}
& = & \beta, \label{dsdu} \\
\left  (  \frac{\partial  S_G}{\partial  W}  \right  )_{U}  &  =  &
\gamma. \label{dsdvar}
\end{eqnarray}
%
\section{Non-additivity}
\ll{nonadd}
%
We remark that  although the thermodynamics of a system  in the EGE is
well  defined by  the equations  in the  previous section,  it  is not
straighforward  to establish  a mutual  equilibrium condition  for two
different  systems that  would allow  to establish  a zeroth  law (or,
equivalently,   an   intensive  temperature)\cite{Hornix1970}.    This
problem  is  due  to  the  non additive  character  of  the  potential
$\Phi(\beta, \gamma)$.  Let us consider  two systems $1$ and  $2$ with
hamiltonians $H_1$  and $H_2$.  By applying the  rules derived  in the
previous sections independently to the two systems, one can derive the
thermodynamic            potentials            $\Phi_1(\beta,\gamma)$,
$\Phi_2(\beta,\gamma)$    as    well     as    the    mean    energies
$U_1(\beta,\gamma)$ and $U_2(\beta,\gamma)$. One can then try to solve
the composite system  $1+2$ with hamiltonian $H_1+H_2$. It  is easy to
verify that the new potential $\Phi_{1+2}(\beta,\gamma)$ satisfies:
\begin{eqnarray}
\nonumber
&& \Phi_{1+2}(\beta,\gamma)=\Phi_{1}(\beta,\gamma)+\Phi_{2}(\beta,\gamma)\\
&& -  \ln \left \langle e^{\gamma 
\left [     (H_1+H_2-U_{1+2})^2 - (H_1-U_1)^2 - (H_2-U_2)^2 \right ] 
} \right \rangle,
\end{eqnarray}
where $U_{1+2}$ is the mean energy of the composite system.  Note that
$U_{1+2}$ as well as the average indicated by the angular brackets are
computed  with  the  probability  distribution  corresponding  to  the
composite  system $1+2$  which, in  general,  cannot be  written as  a
product of probability distributions for systems $1$ and $2$.

The average  values are, in general, non-additive  ($U_{1+2}\neq U_1 +
U_2$). But even if additivity  of $U$ is imposed, the potential $\Phi$
remains non-additive.  The correction  term depends on the microscopic
details of the two hamiltonians $H_1$ and $H_2$.

This  lack of  additivity  does  not allow  to  define an  equivalence
relation  of ``mutual''  equilibrium. Consider  that two  systems are
known to be independently in  equilibrium with a bath characterized by
$\beta$ and $\gamma$.  On putting the two systems in contact with bath
simultaneously,  their   properties  (probability  distribution)  will
change.   The  parameters  $\beta$  and  $\gamma$  are,  consequently,
properties of the  bath and cannot be considered  as properties of the
system, contrary to what occurs  with the corresponding $\beta$ in the
standard canonical ensemble.
%
\section{Stability criterion}
\label{Stabi}
%
In  standard thermodynamics,  the  stability criterion  $-\partial U  /
\partial \beta >0$  is derived from the condition  of maximum entropy.
The derivation \cite{Callen1985} considers  a partition of an isolated
system into  any two  subsystems.  By allowing  the two  subsystems to
alter their energies at fixed  total energy, one can analyze
the entropy  change when  the system is  (virtually) displaced  out of
equilibrium.  The  condition of the  maximum total entropy allows  to deduce
that  the equilibrium state  corresponds to  a state  with homogeneous
$\beta$  (equilibrium  condition)  and  with  $-\partial  U  /\partial
\beta>0$  (stability condition).   For  the derivation,  nevertheless,
additivity of the entropy of the two subsystems must be used.

Within  our new formalism,  an additivity  assumption cannot  be used.
Therefore  it   is  not  straightforward  to   establish  a  stability
criterion.  Although we cannot give  a rigorous proof, in this section
we  provide  some evidences  that  the  same  criterion ($-\partial  U
/\partial \beta>0$) must hold.

First  of  all, it  interesting  to  remark  that the  requirement  of
Eq.~(\ref{defu})  that allows  to  find  $U$ can  be  rewritten as  an
extremal    condition.      Consider    the    definition     of    an
``out-of-equilibrium'' potential:
\begin{equation}
\Psi(\beta,\gamma,U)=\ln Z.
\end{equation}
Note  that  here we  are  considering  $\beta$,  $\gamma$ and  $U$  as
independent  variables.  $U$  shall not  be regarded  as  the internal
energy but  as a  parameter that allows  virtual displacements  out of
equilibrium.    It   can  be   checked   that   the  self   consistent
Eq.~(\ref{defu}) can then be written as:
\begin{equation}
\left ( \frac{\partial \Psi}{\partial U} \right )_{\beta,\gamma} = 0.
\end{equation}
Therefore,  only  the  extrema  of $\Psi$  correspond  to  equilibrium
solutions: $U={\tilde U}(\beta,\gamma)$. By substituting in $\Psi$ one
gets the equilibrium potential:
\begin{equation}
\Phi(\beta,\gamma)=\Psi(\beta,\gamma,{\tilde U}(\beta,\gamma)).
\end{equation}
Secondly,  from  Eq.~(\ref{defzg})  note   that  if  $\gamma  >0$  and
$U\rightarrow \pm \infty$, then $Z_G\rightarrow 0$ and therefore $\Psi
\rightarrow -\infty$.

We can use this ``out-of-equilibrium'' potential to define a stability
criterion.  It is straighforward to compute its second derivative:
\begin{equation}
\left ( \frac{\partial^2  \Psi}{\partial U^2} \right )_{\beta,\gamma}=
2\gamma \left ( 2\gamma W -1 \right ).
\end{equation}
Note  that  the  positivity  of  $\cal C$  in  Eq.~(\ref{defc})  would
guarantee  that  this second  derivative  of  $\Psi$  is negative  and
therefore  the   state  of   equilibrium  corresponds  to   maxima  of
$\Psi(\beta,\gamma,U)$ with  respect to $U$  displacements.  Although,
contrary  to what  happens  in the  standard  canonical ensemble,  the
positivity of $W$ does not  ensure the sign of $-\partial U /
\partial \beta$ in general, at least  we can derive that for small and
positive values of $\gamma$:
\begin{equation}
0< -  \left (  \frac{\partial U}{ \partial  \beta} \right  )_{\gamma }
\Rightarrow  \left  (   \frac{\partial^2  \Psi}{\partial  U^2}  \right
)_{\beta,\gamma} <0.
\end{equation}
Thus we suggest that this is the stability criterion to be used within
the EGE, and we will use it in Section \ref{Appli} for the analysis of
some examples.
%
\section{Comparison with $q$-exponential distributions}
\label{Qexp}
%
$q$-exponential  distributions  are  the  central predictions  of  the
generalized     statistical    mechanics    proposed     by    Tsallis
\cite{Tsallis1988}.  These distributions have been considered as model
distributions to describe various  complex systems at their stationary
states \cite{Abeokamoto2001ed,Kaniadakis2002ed,Gelmann2003ed}.
   The general form of  such distributions is given by $p(x)\sim
e_q(x)$,  where the  $q$-exponential  is  defined as  $e_q(x)  = [1  +
(1-q)x]^{1/(1-q)}$.  This function  goes to the
usual $\exp(x)$ function for $q\to 1$. For definiteness, we restrict to the range $0<q<1$.

In this section, we compare the $q$-exponential distributions with the
equilibrium distributions of the EGE. But first, we show how to derive
the  $q$-exponential  distributions   by  generalizing  the  canonical
ensemble  approach,  along the  lines  of  section \ref{Contact}.   We
define a  parameter which is in  general, a function of  the energy  $E_2$ of
the reservoir 
\be
\beta(E_2) = \frac{dS_2}{dE_2}.
\label{bete2}
\ee
At equilibrium, it attains the value given by Eq.~(\ref{defbeta}).  We
impose that $\beta(E_2)$ satisfies:
\be \frac{d}{d{E}_2}\left(\frac{1}{\beta(E_2)}\right) = Q,
\label{defq}
\ee
where $Q$ is a positive valued constant. From Eqs.~(\ref{defq})
and (\ref{bete2}), we obtain
\be
\frac{d}{d{E}_2} \beta(E_2)   = \frac{d^2 S_2}{d E_2^2}
 = -Q\beta^2 (E_2).
\label{dbde}
\ee
In general, for all integer values of $n$
\be
\frac{d^n S_2}{d E_2^n} = (n-1)! (-Q)^{n-1} \beta^n (E_2).
\ll{dden}
\ee
Now unlike in Eq.~(\ref{expands2}),  if in the expansion of $S(E-E_1)$
around the  equilibrium value $(E-U)$,  we retain derivatives  of $S_2$
upto all orders, then we have
\be 
S_2(E-E_1) = S_2(E-U) + \sum_{n=1}^{\infty}
         \frac{1}{n!} \left . \frac{d^n S_2}{d E_2^n} \right|_{E-U}
          (U-E_1)^n.
\ll{exps2}
\ee
On applying Eq.~(\ref{dden}) for the case of equilibrium,  we can write
\be
S_2(E-E_1) = S_2(E-U) + \sum_{n=1}^{\infty}
           \frac{1}{n} (-Q)^{n-1} \beta^n  (U-E_1)^n,
\ll{exps22}
\ee
where note  that $\beta$  is given by  its value at  equilibrium.  The
equilibrium probability distribution  is then given from (\ref{prob1})
as
\be
p(E_1) \sim \exp\left[\sum_{n=1}^{\infty}
           \frac{1}{n} (-Q)^{n-1} \beta^n  (U-E_1)^n \right].
\ll{psum}
\ee
To compare Eq.~(\ref{psum}) with the $q$-exponential 
distribution given by
\be
p_q(E_1) \sim e_q[\beta(U-E_1)] = [1 + (1-q)\beta(U-E_1)]^{1/(1-q)},
\ll{canonpq}
\ee
we rewrite the $q$-exponential as 
\be e_q[\beta(U-E_1)]  = \exp\left[\frac{ \ln  [1 + (1-q)\beta(U-E_1)]
}{(1-q)} \right], \ll{rewriteq} 
\ee
and  expand  the $\ln$  function  using the  series  $\ln[1+x]  = x  -
\frac{x^2}{2} +\frac{x^3}{3} -\frac{x^4}{4} + ... $, provided that $-1
< x \le 1$. Thus we can write
\be
e_q[\beta(U-E_1)] = \exp\left[\sum_{n=1}^{\infty}
           \frac{1}{n} \{-(1-q)\}^{n-1} \beta^n  (U-E_1)^n \right],
\ll{canonseriesq}
\ee
for  $-1 <  (1-q)\beta(U-E_1) \le  1$.  Thereby,  on identifying  $Q =
(1-q)$  we  may  say  that  the general  equilibrium  distribution  of
Eq.~(\ref{psum}) based on  assumptions (\ref{bete2}) and (\ref{defq}),
is  identical to  a $q$-exponential  distribution.  Assuming  that the
relevant $q$ values  are quite close to unity, we  may keep terms only
upto  second   order  as  done  in   Eq.~(\ref{expands2}).   Then  the
equilibrium $q$-distribution  for system 1 being in  microstate $i$ of
energy $\epsilon_i$ can be written as
\begin{equation}
p_{q}(\epsilon_i)  = \frac{1}{Z_q}{\exp} \left  [ -\beta  \epsilon_i -
\frac{1}{2}(1-q) \beta^2 (\epsilon_i -U)^2 \right ],
\label{pits}
\end{equation}
where $Z_q$ is the normalization constant. 

On  the  other hand,  for  the  case of  EGE,  instead  of fixing  the
derivative of  $\beta^{-1}$ (Eq.~(\ref{defq})), we  fix the derivative
of $\beta$ as follows:
\be
\frac{d}{d{E}_2} \beta(E_2) = -2\gamma,
\ee
where $\gamma$ is  independent of $E_2$. This ensures  that the higher
order ($n>2$) derivatives of  $S_2$ vanish.  On comparing (\ref{pits})
and (\ref{piexp}),  we note that  $(1-q)$ plays the role  analogous to
$\gamma$.

It  may  be remarked  that  if  we  identify parameter  $\beta(E_2)  =
1/T(E_2)$ as the  inverse temperature,  then Eq.~(\ref{defq}) implies
that  the   heat  capacity   of  the  reservoir   $C_2  =   dE_2/dT  =
Q^{-1}$.  Recently,   the  $q$-exponential  distributions   have  been
discussed  in the  context of  a reservoir  with finite  heat capacity
\cite{Almeida2001}. On the other hand, following Gibbs' approach to 
the canonical ensemble, but instead using the $q$-generalized Boltzmann 
entropy,  $q$-exponential distributions were derived 
in Ref. \onlinecite{Aberajagopalepl55}.

\section{Application to a system of independent spins}
\label{Appli}
\subsection{Single Spin}

\begin{figure}[htb]
\begin{center}
\epsfig{file=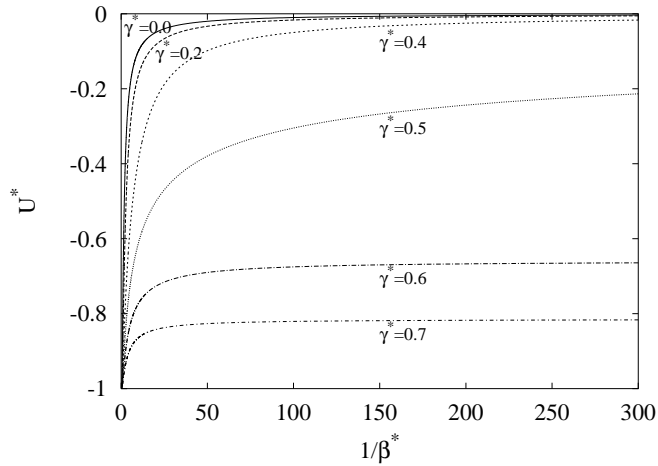,width=9cm}
\end{center}
\caption{\label{FIG1}Behaviour of  the mean reduced energy  $U^*$ as a
function of $1/\beta^{*}$ for several values of $\gamma^*$ in a system
of a single spin.}
\end{figure}

\begin{figure}[htb]
\begin{center}
\epsfig{file=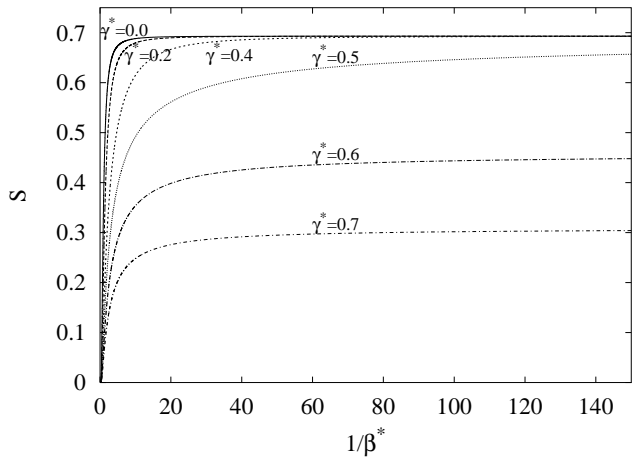,width=9cm}
\end{center}
\caption{\label{FIG2}Behaviour  of the  entropy $S$  as a  function of
$1/\beta^*$ for several  values of $\gamma^*$ in a  system of a single
spin.}
\end{figure}

As a first  example of the EGE, we apply our  formalism 
to  the problem  of a  system  with only  two energy  levels.  Let  us
consider a single spin $s=\pm 1$  in the presence of a constant external 
magnetic field $B$.  The hamiltonian of the system reads:
\begin{equation}
H=-Bs.
\end{equation}
The partition function is given by:
\begin{equation}
Z_G = e^{\beta B} e^{-\gamma (-B-U)^2} +  e^{-\beta B} e^{-\gamma (B-U)^2},
\label{eq1}
\end{equation}
where the mean energy $U$ is the solution of the self-consistent equation:
\begin{equation}
U = -B e^{\beta B} e^{-\gamma (-B-U)^2} +  B e^{-\beta B} e^{-\gamma (B-U)^2}.
\label{eq2}
\end{equation}
The dependence on $B$ can be easily overcome by defining the reduced units:
\begin{equation}
U^* = U/B, \; \; \; \beta^* = \beta B, \; \; \; \gamma^* = \gamma B^2.
\end{equation}
Thus, Eq. (\ref{eq2}) becomes:
\begin{equation}
U^* = e^{-\beta^*}  e^{-\gamma^* (1-U^*)^2} - e^{\beta^*} e^{-\gamma^*
(1+U^*)^2}.
\label{eq12}
\end{equation}
The numerical solution of this equation is plotted in Fig.~\ref{FIG1}.
The behaviour  of $U^*$  as a function  of $1/\beta^{*}$ is  shown for
different values  of $\gamma^*$.   For $\gamma^*=0$, one  recovers the
behaviour $U^*=\tanh(\beta^*)$  corresponding to the case  of a system
in contact with a infinite reservoir.  For $\gamma^* \neq 0$, $U^*$ is
smaller, indicating that it is more dificult to disorder the system by
decreasing   $\beta^*$.   It   is   interesting  to   note  that   for
$\gamma^*\simeq0.5$, there  is a change  in the behaviour  at $\beta^*
\rightarrow 0$. Above this value of $\gamma^*$, the system is not able
to   disorder  completely   anymore   and  always   keeps  a   certain
magnetization ($m=\langle s \rangle =-U^*$). This can be regarded as a
``phase transition'',  that occurs at $\beta^*=0$. This  change in the
behaviour  occuring at  $\gamma^*  \simeq  0.5$ can  also  be seen  by
plotting the  entropy $S$ as  a function of $1/\beta^*$  for different
values  of  $\gamma^*$.   This  is  shown  in  Fig.  \ref{FIG2}.   For
$\gamma^*< 0.5$ the entropy tends  to $\ln 2$ for $\beta^* \rightarrow
\infty$, whereas it tends to a lower value for $\gamma^*>0.5$.
\subsection{Two  spins}
\begin{figure}[htb]
\begin{center}
\epsfig{file=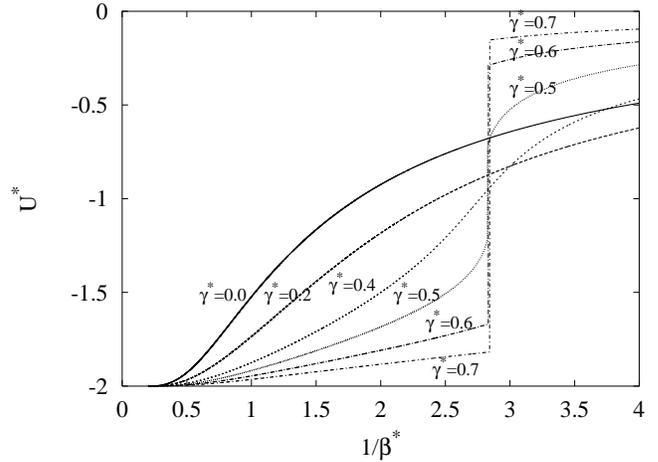,width=9cm}
\end{center}
\caption{\label{FIG3}Behaviour of the mean  energy $U^*$ as a function
of $1/\beta^{*}$ for  several values of $\gamma^*$ in  a system of two
independent spins.}
\end{figure}

\begin{figure}[htb]
\begin{center}
\epsfig{file=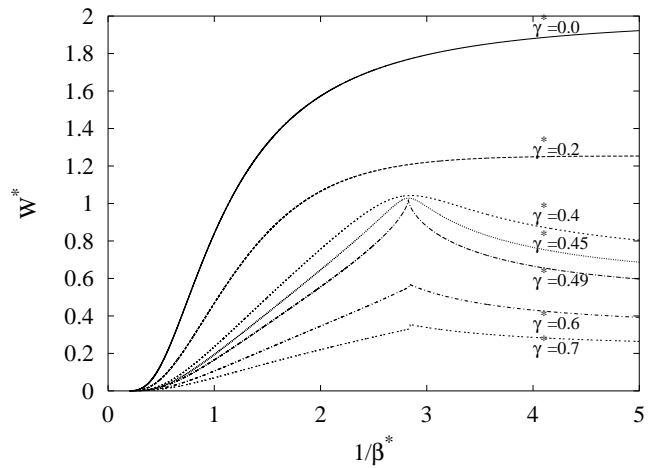,width=9cm}
\end{center}
\caption{\label{FIG4}Behaviour  of   the  energy  fluctuations   as  a
function of $1/\beta^{*}$ for several values of $\gamma^*$ in a system
of two independent spins.}
\end{figure}

\begin{figure}[htb]
\begin{center}
\epsfig{file=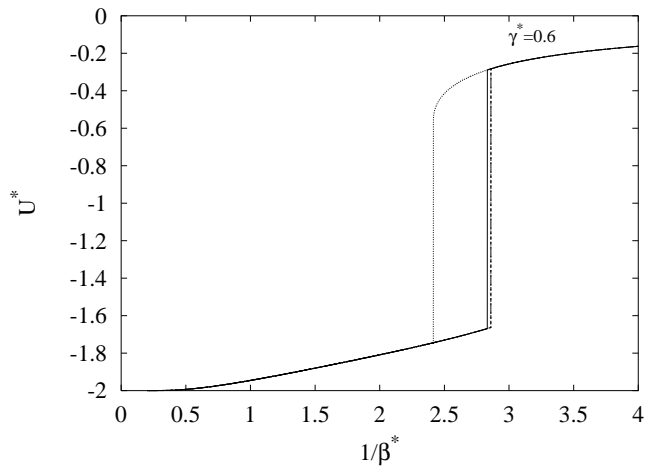,width=9cm}
\end{center}
\caption{\label{FIG5}Behaviour of $U^*$ as a function of $1/\beta^{*}$
for a  system of  two spins with  $\gamma^*=0.6$. The  continuous line
represents  the equilibrium  solution  (with maximum  $\Psi$) and  the
dashed lines represent metastable  solutions which correspond to local
maxima.}
\end{figure}

As a second step, it is also very instructive to study a system of two
independent spins.   This will illustrate  the non-extensive behaviour
of the  solution.  In  this case, the  numerical solution of  the self
consistent  equation  (\ref{defu}) for  the  mean  energy renders  the
behaviour shown in Fig.  \ref{FIG3}.   For the values of $\beta^*$ and
$\gamma^*$ for which more than one solutions are possible, we have used
the  stability criterion  proposed  in Section  \ref{Stabi} to  decide
which is the ``equilibirum'' solution.   As can be seen, for $\gamma^*
> 0.49$ a discontinuity occurs associated  with a sudden loss of order
in  the system.   Although the  system is  far from  the thermodynamic
limit, this  change shares many similarities with  a phase transition.
Fig. \ref{FIG4}  displays the behaviour of  the corresponding energy
fluctuations.   It can  be  seen  that $W^*=  (\langle  H^2 \rangle  -
U^2)/B^2$ exhibits a cusp at the transition for $\gamma^*\simeq 0.49$.
For  larger   values  of   $\gamma^*$,  the  fluctuations   exhibit  a
discontinuity.   The discontinuities  are associated  with first-order
phase transitions  that display metastable behaviour.   As an example,
in  Fig.  \ref{FIG5}, we  show the  detailed behaviour  of $U^*$  as a
function   of   $1/\beta^*$   for   $\gamma^*=0.6$.   In   the   range
$2.42<1/\beta^*<2.86$, the  numerical analysis of  the self-consistent
equation renders three solutions.

\begin{figure}[htb]
\begin{center}
\epsfig{file=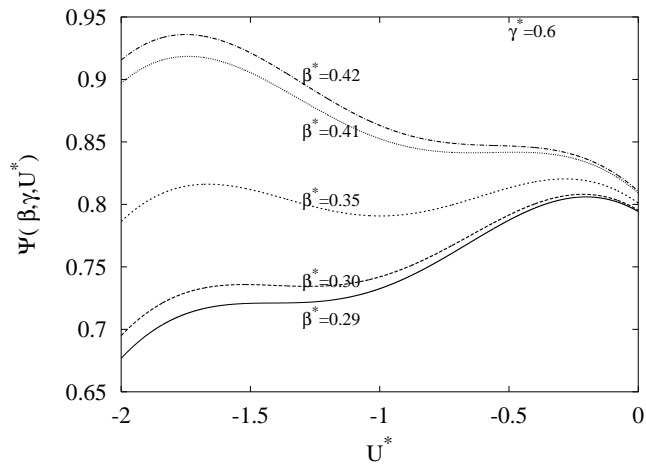,width=9cm}
\end{center}
\caption{\label{FIG6}Behaviour         of         the        potential
$\Psi(\beta^*,\gamma^*,u^*)$ for different values of $1/\beta^*$ for a
system  of two spins  with $\gamma^*=0.6$.   The equilibrium  value of
$U^*$ corresponds to the maximun of the potential $\Psi$.}
\end{figure}

By  analyzing the behaviour  of the  potential $\Psi(\beta^*,\gamma^*,
U^*)$, shown in Fig.  \ref{FIG6} it is easy to verify that two of such
solutions are  stable (correspond to  local maxima of  $\Psi$) whereas
one is unstable  (corresponds to a local minimum of  $\Psi$ and is not
plotted  in Fig.   \ref{FIG5}).   The equilibrium  transition jump  at
$1/\beta^*  \simeq 2.832$  is determined  by the  equality of  the two
maxima of $\Psi$.

\begin{figure}[htb]
\begin{center}
\epsfig{file=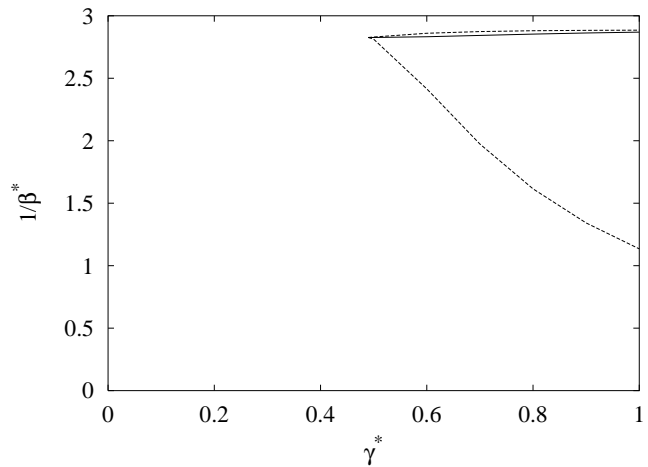,width=9cm}
\end{center}
\caption{\label{FIG7}Phase  diagram for  a system  of two  spins.  The
continuous    line   represents   first-order    transitions   whereas
discontinuous lines indicate the metastability limits.}
\end{figure}

For   the  system   of   two   spins,  therefore,   we   can  plot   a
$\beta^*-\gamma^*$ phase diagram, shown  in Fig.  \ref{FIG7}. The line
of  first order  phase transitions  ends  in a  ``critical'' point  at
$\beta^*  \simeq 0.353$  and $\gamma^*  \simeq 0.49$.   This  point is
characterized by  the condition $1/W^*=2\gamma^*$  and thus, according
to Eq.~(\ref{d2phib}), corresponds  to a divergence of $\cal  C$ but not
to a divergence of the  fluctuations $W^*$, that can never diverge for
such a system with a finite number of bounded energy levels.
\subsection{Several spins}
\begin{figure}[htb]
\begin{center}
\epsfig{file=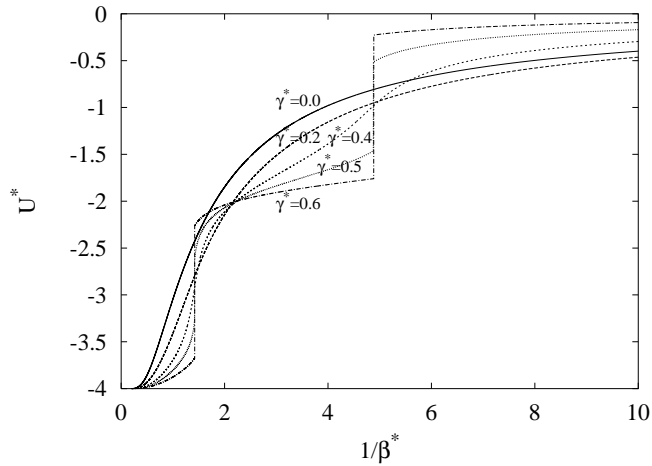,width=9cm}
\end{center}
\caption{\label{FIG8}Behaviour of the mean  energy $U^*$ as a function
of $1/\beta^{*}$ for  several values of $\gamma^*$ in  a system of 4
independent spins.}
\end{figure}

We have also performed a numerical study of systems with larger number
of independent spins in the presence of an external field.  An example
is shown in  Fig.  \ref{FIG8}, corresponding to a  system with 4 spins
(16 energy  levels). A sequence  of two consecutive  phase transitions
can be observed.  As an interesting remark we want to note that in the
case  of $N$  ``non-interacting''  spins $s_k  (k=1,\dots  N)$ in  the
presence of  an external field,  long-range forces will appear  due to
the finite size  of the bath.  This can be easily  seen by writing the
probabilities $p_i$ for the  microstates ($i=1,\cdots, 2^N$) of such a
system:
\begin{equation}
p_i=  \exp\left[ {\beta^*  \sum_{k=1}^N s_k  -  \gamma^*  \left (  \sum_{k=1}^N
s_k-U^* \right )^2}\right ].
\end{equation}
Note that  the development  of the squared  term in the  exponent will
lead to terms $-\gamma^* s_k s_j$ wich correspond to antiferromagnetic
interactions  among all  spin pairs.  A more  detailed study  of these
examples is out of the scope of this paper.
%

\section{Summary}
\label{Concl}
We  have   presented  the  extended  gaussian  ensemble   (EGE)  as  a
generalization  of  the  standard  canonical ensemble.   The  ensemble
statistics have  been derived by  two methods: first by  considering a
system  in contact  with  a  finite bath,  secondly  from the  maximum
statistical entropy principle by fixing the knowledge of both the mean
energy  and the  energy  fluctuations.  The  obtained probability  law
depends on two parameters $\beta$ and $\gamma$ which are properties of
the bath.   Thermodynamic relations have  been derived and  a possible
stability criterion  has been  suggested.  Nevertheless this  point as
well  as  the  possibility   for  establishing  a  mutual  equilibrium
criterion will  need further  analysis in future  works. We  have also
presented an  application of  the EGE formalism  to the analysis  of a
system of one spin and two independent spins.  Among other interesting
results, the most remarkable one is the possibility for occurence of a
critical point or first-order  phase transitions induced by the finite
size  of the  reservoir.  Further,  comparisons of  this  new ensemble
formalism  with   alternative  extensions  of   statistical  mechanics
proposed        for       the        study        of       nanosystems
\cite{Hill1994,Hill2001a,Hill2001b}   or   for   other   non-extensive
systems, are interesting problems for research.

\section{Acknowledgements}
This work  has received financial support from  CICyT (Spain), project
MAT2001-3251 and CIRIT  (Catalonia), project 2000SGR00025.  R.S. Johal
acknowledges financial support from  the Spanish Ministry of Education,
Culture and Sports. The authors acknowledge fruitful discussions with
Dr. Avadh Saxena.

\end{document}